\documentclass{article}

\usepackage[utf8]{inputenc}
\usepackage[T1]{fontenc}
\usepackage[english,naustrian]{babel} 
\usepackage{csquotes}
\usepackage[table,xcdraw]{xcolor}
\usepackage{placeins}
\usepackage{float}
\usepackage{pdfpages}
\usepackage{lscape}

\usepackage{amsmath,amssymb,amsthm,graphicx} 
\usepackage[margin=1in]{geometry} 
\usepackage{subfiles}

\usepackage[
backend=biber,
natbib
]{biblatex}
\bibliography{bibliography}

\usepackage{graphicx}
\graphicspath{{img/}}

\usepackage{tikz}
\usepackage{pgfplots}
\pgfplotsset{compat=1.16}

\usepackage{hyperref}

\usepackage{todonotes}

\usepackage{booktabs}

\usepackage{authblk}
\title{Supervised Machine Learning Classification for Short Straddles on the S\&P500}

\author[1]{Alexander~Brunhuemer}
\author[1]{Lukas~Larcher}
\author[2]{Philipp~Seidl}
\author[1]{\\ Sascha~Desmettre}
\author[2]{Johannes~Kofler}
\author[1]{Gerhard~Larcher}
\affil[1]{JKU Linz -- Institute for Financial Mathematics and Applied Number Theory}
\affil[2]{JKU Linz -- Institute for Machine Learning}

\begin{document}
\selectlanguage{english}

\maketitle

\begin{abstract}
In this working paper we present our current progress in the training of machine learning models to execute short option strategies on the S\&P500. As a first step, this paper is breaking this problem down to a supervised classification task to decide if a short straddle on the S\&P500 should be executed or not on a daily basis. We describe our used framework and present an overview over our evaluation metrics on different classification models. In this preliminary work, using standard machine learning techniques and without hyperparameter search, we find no statistically significant outperformance to a simple ``trade always'' strategy, but gain additional insights on how we could proceed in further experiments.
\end{abstract}

\section{Introduction}
The investigations presented in this working paper are an extension of the papers ``Analysis of Option Trading Strategies Based on the Relation of Implied and Realized S\&P500 Volatilities'' \cite{Brunhuemer.2021}, ``Modeling and Performance of Certain Put-Write Strategies'' \cite{Larcher.2013} and ``A Comparison of Different Families of Put-Write Option Strategies'' \cite{Larcher.2012}. In these papers we analyzed the historical performance of certain short option strategies based on the S\&P500 index between 1990 and 2020. In the latest publication we sought to explain outperformance of such strategies based on relations between the implied and the realized volatility of the underlying by modeling the negative correlation between the S\&P500 and the VIX and Monte Carlo simulation. Our research was also based on previous investigations on systematic overpricing of certain options (see for example \citep{Day.1997}, \citep{Ungar.2009} and \citep{SantaClara.2009}).

One of our tested strategies, the Lambda strategy (or better known as short straddle, but we like to call it Lambda strategy, since it is very descriptive for the payoff function), led to great success in real trading. Therefore, analyzing this strategy and finding not really good performance in our (very static) backtesting was somehow surprising for us. In the paper we stated that we think, more dynamic decision-finding about how to invest in these kind of strategies could prove helpful -- maybe by using some machine learning approaches. This working paper is a first step to following our own suggestion and starts out with breaking the stated problem down into a small approachable classification task. Casually spoken, our goal is to train a machine learning model to decide if one should open a basic (``naked'') contract following the lambda strategy for given market data, or not. 

\section{Machine learning framework}
For our machine learning approach we follow a 6 step framework for machine learning projects, as is for example explained in \cite{ZTM.2020} or \cite{Bourke.2019}, which consists of the following steps:

\begin{enumerate}
    \item \emph{Problem definition:} describes the concrete problem we are trying to solve, which is in our case a description of the trading strategy and the according workflow
    \item \emph{Data:} describes the available data
    \item \emph{Evaluation:} describes measures for the quality of our approaches, and what would be a successful model
    \item \emph{Features:} describes the features we are modelling and which data we actually use for this
    \item \emph{Modelling:} which models are we trying and how do we compare them
    \item \emph{Experiments:} based on our findings from before, we can here decide, which of the previous steps we want to adapt and try out new approaches
\end{enumerate}

We will also follow this structure throughout this paper.

\subsection{Description of the trading strategy and the work flow}
\begin{figure}[H]
      \centering
      \begin{tikzpicture}
        \begin{axis}[
            axis x line=center,
            axis y line=center,
            xlabel={$S(T)$},
            xlabel style={below right},
            xmin=-0.5,
            xmax=3,
            ymin=-1,
            ymax=0.8,
            xtick={1.5},
            ytick=\empty,
            xticklabels={$S(0)$},
            yticklabels=\empty,
            legend pos=south east]
            \addplot[blue,mark=none,domain=0:1.5]{0.5-max(0, 1.5-x)};
            \addplot[blue,mark=none,domain=1.5:2.9]{0.5-max(0,x-1.5)};
        \end{axis}
      \end{tikzpicture}
      \caption{Profit function of a pure lambda (short straddle) without securing long positions at the money. $S$ is the S\&P500 index value. The profit/loss (above/below the horizontal axis) depends on the final value at time $T$.}
      \label{fig:lambdaProfit}
  \end{figure}
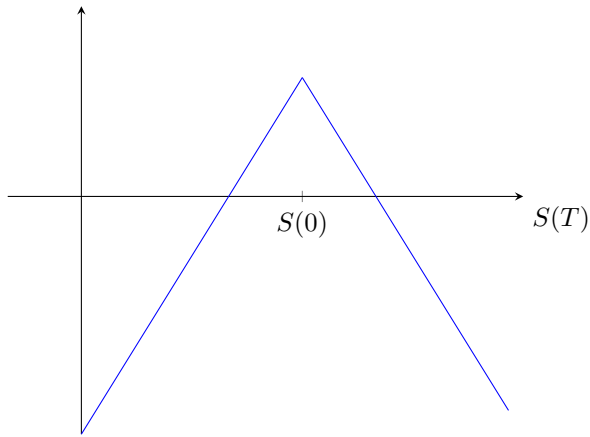

Our machine learning investigation is aiming at optimizing the Lambda strategy, which is in principle based on selling both call- and put options at-the-money. The model should in a first step be able to decide for given market data, if the strategy (i.e. selling one put and one call at the same strike) should either be executed at-the-money, or not. By looking at the basic structure (see Fig. \ref{fig:lambdaProfit}) we see that calm markets with little volatility would work best to keep the gained premium. We could open long positions to limit losses, or one could react to changing market environments by trading the underlying asset (or futures of it for keeping trading costs low), or close the open positions if a certain threshold for losses is reached. These adaptations will not be followed in this paper, but should be kept in mind for further research. The strict rules for a general form of such Lambda strategies are as follows:

\begin{itemize}
\renewcommand\labelitemi{--}
\item We choose a fixed time period of length $T$ (e.g. 2 months, 1 month, one week, two trading days, \dots)

\item For a given day $t$ we trade SPX options with remaining time to expiration $T$ (or with the shortest possible time to expiration larger than or equal to $T$ and with new trading upon expiration of these options). We assume to initiate the trade always at the close time of the trading day $t$.

\item We always go short on the same quantity of call- and put options with same time to expiration (approximately) $T$ and strike $K_1$ as close at-the-money as possible (i.e., with a strike as close as possible to the current value of the S\&P500). 

\item In the case, where we aim to limit losses, we go long the same quantity of put options with the same expiration and with a strike $K_2 < K_1$, and/or we go long on the same quantity of call options with the same expiration and with a strike $K_3 > K_1$.

\item Thus, upon entering the trade, we receive a positive premium of $M$ USD, which is given by the price of the short positions minus the price of the long positions.

\item Our reference currency in all cases is the U.S. dollar (USD).

\item For training our machine learning models we assume to always trade one contract of options. When actually executing these trades one would decide on the number of options to trade, by determining the required margin and calculate the possible number of contracts with the available capital.

\item In some of the strategies the positions are held until expiration. Some of the strategies are equipped with an exit strategy, which means: All contracts are closed as soon as the losses from the call and put positions (since the last trading day) exceed a certain pre-defined level.

\item Now the strikes $K_2$ and/or $K_3$ of the long positions are chosen on the basis of various parameters (depending on what strategy we are looking at). They will always depend on the current value of the S\&P500 (at the trading date); in some cases they will also depend on the value of the VIX or on a certain historical volatility, while in other cases they will depend on the prices of the put and/or call options in question.

\item The trading assumptions in each case (bid/ask prices, the exact trading time, transaction costs, setting of a ``buffer'') are described in the section discussing the backtests of the Lambda strategies. 
\end{itemize}

In this initial approach we ignore the possibility to buy long positions at strikes $K_2$ and $K_3$ and only sell the options at $K_1$. Our machine learning problem now concretely tries to decide for a given trading day (and respective given market data) if the strategy should be executed or not. We assume that the decision is made at market close time on a daily basis.

\FloatBarrier

\subsection{Data}
In our initial approaches to this machine learning problem we restrict ourselves to the following available data:
\begin{itemize}
    \item daily historical put- and call option price data\footnote{we obtained them from the CBOE data shop (\url{https://datashop.cboe.com/})}, which includes amongst others:
    \begin{itemize}
        \item last ask and -bid prices for any available strikes and expiry date
        \item open, close, high and low prices per day
        \item traded volume and open interest
    \end{itemize}
    \item daily publicly available market data, such as
    \begin{itemize}
        \item close, open, high and low of the underlying
        \item close, open, high and low of derived products, such as the VIX for the S\&P500
        \item interest rates
    \end{itemize}
\end{itemize}

The historical option price data would be available to us also for earlier periods, however, only starting with November 2011 the frequency of offered option expiration dates increased (because of the introduction of the SPXW options).

The data is clearly structured, and is available on a very consistent basis. Typically, the data can be gathered on a streaming basis, a monthly interval worked out very well for us in the past. Also, e.g. the CBOE datashop mentioned above provides data on a monthly basis, so this would work as a suitable interval for fetching the data.

To that effect, the experiments in this working paper are based on a set of 1941 data samples, gathered and preprocessed from historical data between November 2011 and August 2020.

\subsection{Evaluation criteria}
For the evaluation of the suitability of our trained model to our problem -- and also for comparing different models -- we need some sort of quality measure. Since we are looking at a classification task, the usual metrics for classifiers are an obvious choice for such metrics. However, even if our classifier would be ``really good'' in terms of classifier metrics, it could still end up very bad in terms of profit made, if the classifier misses the most important choices (i.e. when the classifier would invest in the big losses, or not invest in the big gains). Thus, we also consider a second type of metrics, which are all connected to some profit measure.

\subsubsection*{Classification metrics}
For each of our validation and test sets we evaluate the following usual classification metrics automatically. We are using the implementations from the Python-package ``scikit-learn'' and therefore are following their documentation \cite{SciKit.1} for the definitions.

\begin{itemize}
    \renewcommand\labelitemi{--}
    \item \emph{accuracy:} computes the accuracy (default is fraction) of correct predictions
    \item \emph{recall:} recall describes the quality of a classifier to find all positive samples and is given by the ratio
    \begin{equation*}
        \frac{tp}{tp + fn}
    \end{equation*}
    where $tp$ are the true positives and $fn$ are the false negatives
    \item \emph{balanced accuracy:} gives the accuracy of a classifier, adjusted by the probability of the outcome of each class. More precisely, it is defined as the average recall obtained on each class.
    \item \emph{precision:} describes the ability of the classifier to avoid false positives ($fp$) and is calculated by the ratio
    \begin{equation*}
        \frac{tp}{tp + fp}
    \end{equation*}
    \item \emph{average precision:} Precision and recall are two measures, which you cannot improve without worsening the other. You always have to make trade offs in the optimization of these two metrics. For this reason, the precision-recall curve is a very interesting visualization. The average precision metric works as a summarization of the precision-recall curve into one single metric as the weighted mean of precisions at given thresholds $P_n$, where the weights are given by the increase in the recall metric from the previous threshold $(R_n - R_{n-1})$:
    \begin{equation*}
        AP = \sum_n (R_n - R_{n-1}) P_n
    \end{equation*}
    \item \emph{PRC:} the precision recall curve gives the precision-recall pairs for varying thresholds.
    \item \emph{PRC (auc):} collects information of the PRC in one metric by calculating the area under the curve.
    \item \emph{F1 score:} again a combined measure of precision and recall, which can be interpreted as the harmonic mean of these two metrics.
    \begin{equation*}
        F1 = 2 * \frac{recall \cdot precision}{recall + precision}
    \end{equation*}
    \item \emph{Brier score loss:} measures the mean squared difference between predicted probability and actual outcome
    \item \emph{cross-entropy loss:} the loss function used in logistic regression for a classifier which gives a prediction probability $\hat{y}$ to an actual outcome $y$. In the binary case (with $y \in \{0,1\}$ and $p$ the probability of $y=1$) this leads to:
    \begin{equation*}
        L_{\log}(y,p) = -(y \log p + (1-y) \log (1 - p) )
    \end{equation*}
    \item \emph{ROC curve:} the ROC (receiver operating characteristic) curve is determined by plotting the fraction of true positives to the fraction of false positives for varying threshold.
    \item \emph{ROC (auc):} Collects information of the ROC curve in one metric by calculating the area under the curve.
\end{itemize}

\subsubsection*{Profit metrics}
The second class of metrics we are interested in, are metrics corresponding to some profit calculations. First and foremost, we allow all the above ``standard'' classification metrics to be weighted with respect to the corresponding profit made. That means, for a given sample $y$ and a given prediction of our model $\hat{y}$, we are weighting the above metrics with respect to the profit (or loss), one would have achieved with this specific trade. This implies that trades with big gains or losses are weighted more than trades with minimal gain or losses. Additionally, we calculate the following metrics:

\begin{itemize}
    \renewcommand\labelitemi{--}
    \item \emph{total profit:} For given predictions $\hat{y}_i \in \{0=\text{do not trade}, 1 = \text{trade}\}$ and given profits $p_i$, we calculate the total profit by simply calculating the sum:
    \begin{equation*}
        \text{total profit} = \sum^n_i \hat{y}_i p_i
    \end{equation*}
    \item \emph{average profit:} is determined by taking the mean analogously to the total profit above:
    \begin{equation*}
        \text{average profit} = \frac1n \sum^n_i \hat{y}_i p_i
    \end{equation*}
    \item \emph{average profit per trade:} is determined by taking the mean as above, but only where $\hat{y}_i$ is not $0$.
    \item \emph{standard deviation of profit per trade:} is determined by taking the standard deviation of the profits where $\hat{y}_i$ is not $0$.
    \item \emph{downside deviation of profit per trade:} is determined by taking the standard deviation of the profits where $\hat{y}_i$ is not $0$ and $p_i < 0$.
    \item \emph{average fraction of trades:} is calculated by taking the mean of all $\hat{y}_i$, which gives the number of executed trades as a fraction.
\end{itemize}

Since it is naturally our goal for our models to especially predict the correct outcome for trades which lead to big gains or losses, we put our focus into metrics which take the profit into account.

\subsection{Features}
For a given trading day, where we decide about the execution of the strategy, we use the following features:
\begin{itemize}
    \renewcommand\labelitemi{--}
    \item \emph{put price}: we use the average of the last bid- and ask price and reduce it by USD 0.1 for our sell price
    \item \emph{call price}: is determined analogously to the put price
    \item \emph{strike}: current strike price, which is the closest strike price to the current S\&P500 value
    \item \emph{days to expiry}: the number of days to expiration of the options
    \item \emph{S\&P500 close of last 5 trading days relative to current S\&P500 value}: we use the relative values to the current underlying value, since we want the model to use the development of the last trading days in relation, rather than the absolute S\&P500 values.
    \item \emph{VIX close of trading day and the previous 5 trading days}
\end{itemize}

\subsection{Modelling}

We are using the model implementations of the Sklearn Python library for our experiments. In the following, we describe the evaluated models briefly while closely following the descriptions found in the SciKit-Learn documentation \cite{SciKit.1}.

\subsubsection*{Random Forest Classifier}
A Random Forest is an averaging algorithm based on randomized decision trees. It is a perturb-and-combine technique specifically designed for trees. In this sense, a diverse set of classifiers is created by introducing randomness in the classifier construction. The prediction of the ensemble is given as the averaged prediction of the individual classifiers. Each tree in the ensemble is built from a sample drawn with replacement from the training set. Furthermore, when splitting each node during the construction of a tree, the best split is - in our case - found from all input features.

The purpose of these two sources of randomness is to decrease the variance of the forest estimator. Indeed, individual decision trees typically exhibit high variance and tend to overfit. The injected randomness in forests yield decision trees with somewhat decoupled prediction errors. By taking an average of those predictions, some errors can cancel out. Random forests achieve a reduced variance by combining diverse trees, sometimes at the cost of a slight increase in bias. In practice the variance reduction is often significant hence yielding an overall better model.

The scikit-learn implementation combines classifiers by averaging their probabilistic prediction, instead of letting each classifier vote for a single class.

Parameters: \verb|n_estimators=701| The number of trees in the forest.

\subsubsection*{Logistic Regression}
Logistic regression is a linear model, where the probabilities describing the possible outcomes of a single trial are modeled using a logistic function.

We use binary class $l_2$ penalized logistic regression, which as an optimization problem, minimizes the following cost function:

\[\min_{w, c} \frac{1}{2}w^T w + C \sum_{i=1}^n \log(\exp(- y_i (X_i^T w + c)) + 1) .\]

where $w$ are the parameters, $X_i$ the features, $y_i$ the targets and $C$ is the regularization parameter.

We use the “lbfgs” solver, which is an optimization algorithm that approximates the Broyden–Fletcher– Goldfarb–Shanno algorithm, which belongs to quasi-Newton methods. It is especially suitable for small data-sets and very robust.

Additional parameter: \verb|warm_start=True|

\subsubsection*{$k$ Nearest Neighbors (kNN) Classifier}
Neighbors-based classification is a type of instance-based learning or non-generalizing learning: it does not attempt to construct a general internal model, but simply stores instances of the training data. Classification is computed from a simple majority vote of the nearest neighbors of each point: a query point is assigned the data class which has the most representatives within the nearest neighbors of the point.

The type of classifier we use is based on the $k$ nearest neighbors of each query point, where $k$ is an integer value.

For the distance metric for the tree we use two different configurations: euclidian and cosine.

Also, we use two different weight functions in prediction:
\begin{itemize}
    \renewcommand\labelitemi{--}
    \item Uniform:  All points in each neighborhood are weighted equally. For this configuration we set $k = 13$.
    \item Distance: Points are weighted by the inverse of their distance. in this case, closer neighbors of a query point will have a greater influence than neighbors which are further away. Here we use $k = 101$.
\end{itemize}

\subsubsection*{Multi-layer Perceptron Classifier}
A Multi-layer Perceptron (MLP) learns a function $f(\cdot): \mathcal{R}^m \rightarrow \mathcal{R}^o$ by training on a dataset, where $m$ is the number of dimensions for input and $o$ is the number of dimensions for output - which is 1 in our case. Given a set of features $X = {x_1, x_2, ..., x_m}$ and a target $y$, it can learn a non-linear function approximator for classification. It is different from logistic regression, in that between the input and the output layer, there can be one or more non-linear layers, called hidden layers.

\begin{figure}[htp]
\centering
\includegraphics[width=0.40\textwidth]{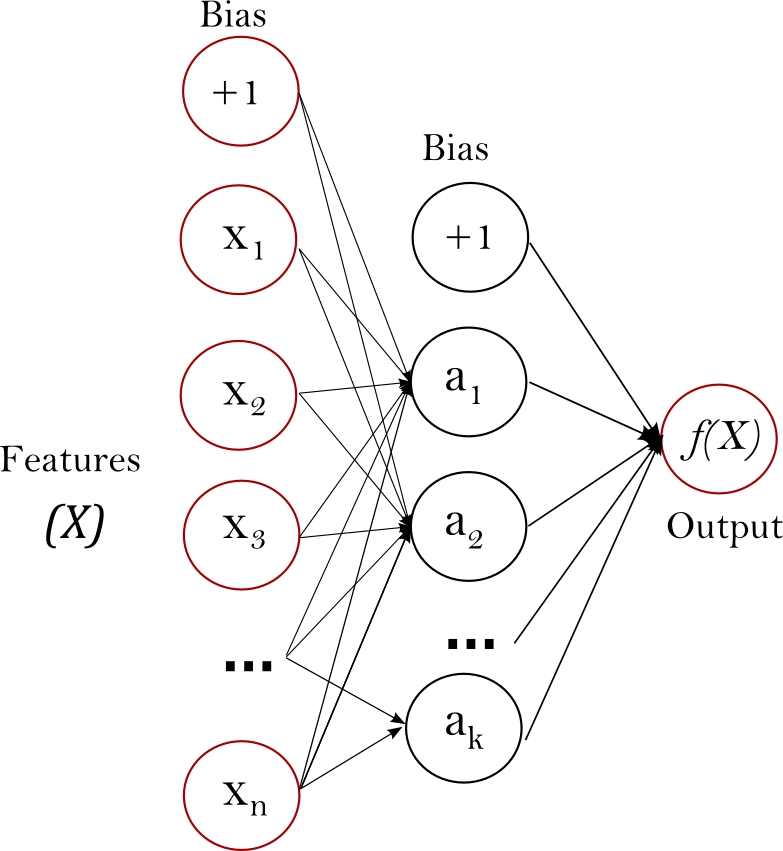}
\caption{One hidden layer MLP (SciKit-Learn \cite{SciKit.2})}
\label{fig:multilayerperceptronNetwork}
\end{figure}

In our case, we use a neural network with two hidden layers of 128 nodes each. We configured the net to use 'relu' as the activation function and, as already with the logistic regression above, 'lbfgs' as the solver.

\subsubsection*{AdaBoost Classifier}
The core principle of AdaBoost is to fit a sequence of weak learners (i.e., models that are only slightly better than random guessing, such as small decision trees) on repeatedly modified versions of the data. The predictions from all of them are then combined through a weighted majority vote (or sum) to produce the final prediction. The data modifications at each so-called boosting iteration consist of applying weights $w_1, w_2, ... w_N$ to each of the training samples. Initially, those weights are all set to $w_i = 1/N$, so that the first step simply trains a weak learner on the original data. For each successive iteration, the sample weights are individually modified and the learning algorithm is reapplied to the reweighted data. At a given step, those training examples that were incorrectly predicted by the boosted model induced at the previous step have their weights increased, whereas the weights are decreased for those that were predicted correctly. As iterations proceed, examples that are difficult to predict receive ever-increasing influence. Each subsequent weak learner is thereby forced to concentrate on the examples that are missed by the previous ones in the sequence.

In our configuration, we use the SAMME.R real boosting algorithm.

\subsubsection*{Gradient Boosting Classifier}
Gradient Tree Boosting is a generalization of boosting to arbitrary differentiable loss functions. It builds an additive model in a forward stage-wise fashion; it allows for the optimization of arbitrary differentiable loss functions. In each stage \verb|n_classes_| regression trees are fit on the negative gradient of the binomial or multinomial deviance loss function. Binary classification as we use it is a special case where only a single regression tree is induced.

Additional parameters: \verb|n_estimators=701, learning_rate=0.5|

\subsubsection*{C-Support Vector Classification}
Given a set of training examples, each marked as belonging to one of two categories, a Support-vector Machine (SVM) training algorithm builds a model that assigns new examples to one category or the other, making it a non-probabilistic binary linear classifier. SVM maps training examples to points in space so as to maximise the width of the gap between the two categories. New examples are then mapped into that same space and predicted to belong to a category based on which side of the gap they fall.

The Sklearn implementation accepts a set of different kernel types to be used in the algorithm. We focused on the RBF kernel with the function $\exp(-\gamma \|x-x'\|^2)$ where we use \verb|1 / (n_features * X.var())| as value of $\gamma$. In our model, we set the parameter $C = 1$.

\subsection{Experiments}

For all our experiments we choose a prequential evaluation approach, as such an approach is very useful for streaming data (e.g. see \cite{Gama.2013}). This means, for given data in a time span $[0,T]$ we split the whole time span into sub-intervals of a given duration $\Delta t$ (this could for example be one month, if new data comes in monthly intervals). Thus, we have $t_i$ in our time span $\{t_0 = 0, t_1, t_2, \dots, t_n = T\}$ such that $t_i - t_{i-1} = \Delta t$.

For a fixed point $t_i$ with $0 < i < (n-1)$ we now split our data into three separate sets. All data available in the time span $[0, t_i]$ are the training set for this iteration, $(t_i, t_{i+1}]$ are the validation set and $(t_{i+1}, t_{i + 2}]$ are the test set for our machine learning models. This means, we train our data on all the data available up to $t_i$, and use the next intervals in time for validation and testing respectively. In the next iteration the training set is extended until $t_{i+1}$ and validation and testing is executed on the subsequent sets, and so forth.

After each iteration, a classification threshold optimization is performed. That is, the algorithm determines above which exact threshold (in steps of 0.1) of the probability given by the model's predictions on the validation set, trading should be performed in order to yield the highest possible average profit.

Thus, we finally get a series of metrics on test sets, which are streaming over time intervals $(t_i, t_{i+1}]$. Based on these, we can either have a look on the behaviour of the metrics over time, or we can calculate statistics (e.g. mean or standard deviations) over a range of such test sets. The exact features and parameters of the executed experiments are given in the following listings of the experiments.

\begin{figure}[htp]
\centering
\includegraphics[width=0.75\textwidth]{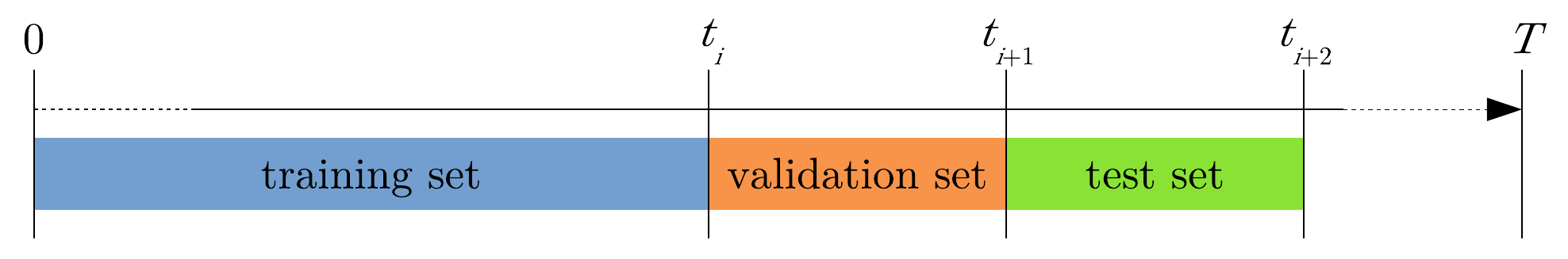}
\caption{Prequential evaluation scheme}
\label{fig:evaluation_timeline}
\end{figure}

\subsubsection*{Experiment 1.1 -- trading on each possible day with very basic data on 1 month streaming data}
\begin{itemize}
    \renewcommand\labelitemi{--}
    \item Iterations: 5
    \item Feature columns: \emph{putPrice, callPrice, strike, spx1, spx2, spx3, spx4, spx5, vix0, vix1, vix2, vix3, vix4, vix5, daysToExpiry}
    \item Prequential split frequency: \emph{1 month}
    \item Start date for test sets: \emph{Feb 2014}
    \item Start date of training set: 2011
    \item Epochs: \emph{10}
    \item Evaluate every n epochs: \emph{1}
\end{itemize}

\subsubsection*{Experiment 1.2 -- trading on each possible day with very basic data on 3 month streaming data}
\begin{itemize}
    \renewcommand\labelitemi{--}
    \item Iterations: 5
    \item Feature columns: \emph{putPrice, callPrice, strike, spx1, spx2, spx3, spx4, spx5, vix0, vix1, vix2, vix3, vix4, vix5, daysToExpiry}
    \item Prequential split frequency: \emph{3 months}
    \item Start date for test sets: \emph{Feb 2014}
    \item Start date of training set: 2011
    \item Epochs: \emph{10}
    \item Evaluate every n epochs: \emph{1}
\end{itemize}

\subsubsection*{Experiment 2.1 -- trading on each possible day with additional data on 3 month streaming data}
\begin{itemize}
    \renewcommand\labelitemi{--}
    \item Iterations: 5
    \item Feature columns: \emph{putPrice, callPrice, strike, spx1, spx2, spx3, spx4, spx5, vix0, vix1, vix2, vix3, vix4, vix5, daysToExpiry, spxHigh, spxLow, vixHigh, vixLow, pmSettled, daysToExpiry}
    \item Prequential split frequency: \emph{3 months}
    \item Start date for test sets: \emph{2014-02}
    \item Start date of training set: 2011
    \item Epochs: \emph{10}
    \item Evaluate every n epochs: \emph{1}
\end{itemize}

\subsubsection*{Experiment 2.2 -- trading on each possible day with additional data on 1 month streaming data}
\begin{itemize}
    \renewcommand\labelitemi{--}
    \item Iterations: 5
    \item Feature columns: \emph{putPrice, callPrice, strike, spx1, spx2, spx3, spx4, spx5, vix0, vix1, vix2, vix3, vix4, vix5, daysToExpiry, spxHigh, spxLow, vixHigh, vixLow, pmSettled, daysToExpiry}
    \item Prequential split frequency: \emph{1 month}
    \item Start date for test sets: \emph{2014-02}
    \item Start date of training set: 2011
    \item Epochs: \emph{10}
    \item Evaluate every n epochs: \emph{1}
\end{itemize}

\subsection{Result overview}
On the following pages we illustrate extractions of the obtained results, based on some exemplary models and in comparison we added the simple ``trade always'' strategy, which is denoted by ``All'' in the metrics table and the visualizations. For each experiment we show two illustration pages, the first page shows (from upper left to lower right)
\begin{itemize}
    \renewcommand\labelitemi{--}
    \item the cumulative profit on the streaming test sets,
    \item the profit per test set with trading always as a baseline (this is not cumulative),
    \item the violin plot on the profits per test set,
    \item the box plot of the profits per test set.
\end{itemize}
On the second page we illustrate four box plots on
\begin{itemize}
    \renewcommand\labelitemi{--}
    \item average precision (upper left) and average precision since 2019 (upper right)
    \item balanced accuracy (lower left) and balanced accuracy since 2019 (lower right)
\end{itemize}

The p-values in the box plots are determined with Wilcoxon tests while applying Bonferroni correction. In addition to these illustrations we provide means of metrics on the test set over all test sets and the test sets since 2019 for each experiment in the appendix. Tendencies which can be deduced from our evaluations are:
\begin{itemize}
    \renewcommand\labelitemi{--}
    \item The Gradient Boost algorithm worked best in terms of total cumulative profit in all experiments.
    \item The violin plot and box plot on the profit show no clear deviation compared to the ``trade always'' strategy. Only the Gradient Boost algorithm in the 1-month split intervals indicate less fat tails, however, nothing is statistically significant.
    \item Trading on 3-month split intervals reduces the profit made in comparison to 1-month split intervals.
    \item The average precision metric is in almost all cases better than in the ``trade always'' strategy, here even statistical significance is reached for some cases, however, this does not translate to actual profit metrics.
    \item The average precision is again better in 1-month split intervals than in 3-month split intervals.
    \item The balanced accuracy became better since 2019 in experiments 2.1 and 2.2, but overall there is no improvement compared to the ``trade always'' strategy in all cases.
    \item The mean average of the number of trades (see last line in the metrics tables in the appendix) indicates, that support vector classifiers tend to trade much more often than other classifiers. AdaBoost classifier is on the low end on this metric throughout all experiments, and Gradient Boost seems to adapt, since the average trades go down drastically in the evaluation since 2019.
\end{itemize}

\includepdf[pages=-,landscape=true]{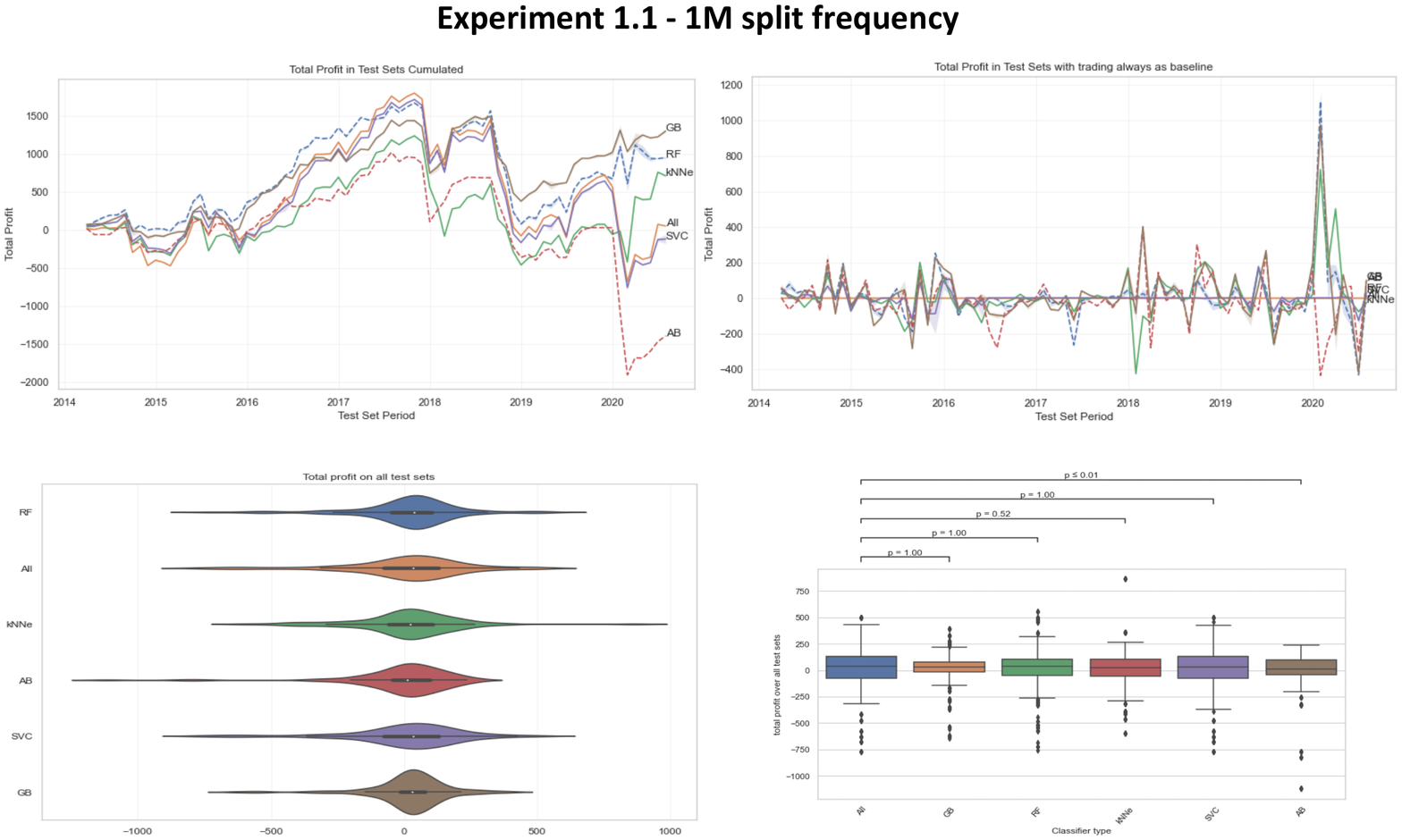}

\subsection{Illustration of a sample predictions timeline}

In this section we want to demonstrate how the predictions made by one of our machine learning models look like over the course of a year. In this sample we use the predictions made between March 1, 2019 and February 28, 2020 by the Random Forest algorithm of Experiment 2.1 as defined above. Trading is considered every Friday and holiday weeks are omitted for simplicity. The values in the ``Prediction'' column are the probabilities determined by our model that entering the Lambda strategy for the following week is superior to not trading at all. The according rows are colored green if the probability is greater than 0.5 (``trade!'') and red otherwise (``don't trade!'').

\begin{table}[h]
    \centering
\ttfamily
\begin{tabular}{ |c|c| } 
 \hline
 Week & Prediction \\
 \hline
 \rowcolor{green}
Week 1 & 0.53780 \\ 
\rowcolor{green}
Week 2 & 0.70899 \\ 
\rowcolor{green}
Week 3 & 0.68474 \\ 
\rowcolor{green}
Week 4 & 0.58345 \\ 
\rowcolor{green}
Week 5 & 0.75892 \\ 
\rowcolor{green}
Week 6 & 0.83024 \\ 
\rowcolor{green}
Week 7 & 0.68759 \\ 
\rowcolor{green}
Week 8 & 0.69330 \\ 
\rowcolor{green}
Week 9 & 0.50499 \\ 
\rowcolor{red}
Week 10 & 0.44936 \\ 
\rowcolor{red}
Week 11 & 0.41084 \\ 
\rowcolor{red}
Week 12 & 0.36519 \\ 
\rowcolor{red}
Week 13 & 0.43224 \\ 
\rowcolor{red}
Week 14 & 0.49786 \\ 
\rowcolor{red}
Week 15 & 0.40942 \\ 
\rowcolor{red}
Week 16 & 0.25678 \\ 
\rowcolor{green}
Week 17 & 0.51641 \\ 
\rowcolor{green}
Week 18 & 0.62767 \\ 
\rowcolor{green}
Week 19 & 0.67760 \\ 
\rowcolor{green}
Week 20 & 0.69900 \\ 
\rowcolor{green}
Week 21 & 0.58345 \\ 
\rowcolor{red}
Week 22 & 0.36091 \\ 
\rowcolor{red}
Week 23 & 0.31954 \\ 
\rowcolor{red}
Week 24 & 0.31954 \\ 
 \hline
\end{tabular}
\quad
\begin{tabular}{ |c|c| } 
 \hline
 Week & Prediction \\
 \hline
 \rowcolor{red}
Week 25 & 0.34522 \\ 
\rowcolor{red}
Week 26 & 0.36519 \\ 
\rowcolor{green}
Week 27 & 0.56205 \\ 
\rowcolor{red}
Week 28 & 0.36805 \\ 
\rowcolor{red}
Week 29 & 0.48930 \\ 
\rowcolor{red}
Week 30 & 0.46505 \\ 
\rowcolor{green}
Week 31 & 0.55350 \\ 
\rowcolor{green}
Week 32 & 0.62767 \\ 
\rowcolor{green}
Week 33 & 0.61769 \\ 
\rowcolor{red}
Week 34 & 0.40514 \\ 
\rowcolor{red}
Week 35 & 0.36519 \\ 
\rowcolor{red}
Week 36 & 0.32240 \\ 
\rowcolor{green}
Week 37 & 0.57061 \\ 
\rowcolor{green}
Week 38 & 0.54208 \\ 
\rowcolor{red}
Week 39 & 0.48645 \\ 
\rowcolor{red}
Week 40 & 0.32240 \\ 
\rowcolor{green}
Week 41 & 0.57489 \\ 
\rowcolor{green}
Week 42 & 0.54351 \\ 
\rowcolor{green}
Week 43 & 0.58773 \\ 
\rowcolor{green}
Week 44 & 0.56491 \\ 
\rowcolor{green}
Week 45 & 0.59058 \\ 
\rowcolor{red}
Week 46 & 0.47504 \\ 
\rowcolor{green}
Week 47 & 0.60200 \\ 
\rowcolor{green}
Week 48 & 0.50927 \\
 \hline
\end{tabular}
\rmfamily
    \caption{Predictions timeline Random Forest 2019-2020}
    \label{tab:predictions_timeline}
\end{table}

\section{Further experiments and open problems}
After building this initial modelling framework and our first tests we are now proceeding with various adaptations of the above experiments. A natural extension would be the inclusion of additional features. This could, on the one hand, be more previous data of the S\&P500 and VIX, and on the other hand more market data as e.g. interest rates, the individual stocks of the S\&P500, commodities, and many more.

However, the currently defined problem might also be too simplistic to actually be profitable. As also mentioned in the preceding paper \cite{Brunhuemer.2021} there exist various adaptations of the pure ``naked'' Lambda strategy. Thus, instead of deciding to execute one such contract based on the pure Lambda strategy, it might be better to ask our model, which of the following variations of the strategy we should execute:
\begin{itemize}
    \renewcommand\labelitemi{--}
    \item $V_0$: trading naked short positions at $K_1$ and hold until expiration
    \item $V_1$: trading naked short positions and close if a certain loss threshold would be reached (losses wrt. the opening of the positions)
    \item $V_2$: trading short positions at $K_1$ and additionally a long put position at $K_2$
    \item $V_3$: trading short positions at $K_1$ and additionally a long call at $K_3$
    \item $V_4$: trading short positions at $K_1$ and additionally long positions at $K_2$ and $K_3$ respectively
    \item $V_5$: trading short positions at $K_1$ and use futures to cover for losses when a certain underlying threshold is reached
    \item $V_6$: do not trade at all
\end{itemize}
It would also be a viable option to decide if it would even be better to execute another option strategy, e.g. if the put-write strategy should be executed instead of the Lambda strategy. But this calls for a thorough further investigation.

Another very interesting approach could be based upon our investigations in our preceding paper about the deviations between implied volatility found in S\&P500 option markets and the actually realized volatility. Instead of directly training the model on executing the Lambda strategy we could train models to estimate the subsequently realized volatility (see for example \cite{Osterrieder.2020} or \cite{Carr.2019}) and trade based upon differences to current implied volatility in the options market. One advantage of this approach would be the possibility to build upon existing research in the estimation of volatility via machine learning models.

And not least, there are a lot of possibilities on the technical side to be explored. That is for instance the systematic optimization of our hyperparameters which we are convinced is essential to tap the full potential of the models. Furthermore we by far haven't reached the limits in terms of the algorithms we use and believe that the application of more complex and modern methods to our problem could yield further insights. In this regard we especially view recurrent neural networks (RNNs) and Hopfield networks (see also \cite{Seidl.2020}) as potentially valuable candidates for our research. It will also be inevitable to continue exploring ways to measure and assess the success of our models, which is why we plan to introduce further metrics such as the Sharpe and Sortino ratios.

\section*{Acknowledgements} \label{sec:acknowledge}
The authors are supported by the Austrian Science Fund (FWF), Project F5507-N26, which is part of the Special Research Program Quasi-Monte Carlo Methods: Theory and Applications, and by the Land Upper Austria research funding.

\printbibliography[heading=bibintoc]
  
\appendix

\begin{landscape}
\section{Metrics of experiment 1.1}
\begin{table}[htp]
\tiny
\begin{tabular}{lllllllllllll}
\hline
\textbf{}                                           & \textbf{AB} & \textbf{All} & \textbf{GB} & \textbf{RF} & \textbf{SVC} & \textbf{kNNe} & \textbf{AB (2019)} & \textbf{All (2019)} & \textbf{GB (2019)} & \textbf{RF (2019)} & \textbf{SVC (2019)} & \textbf{kNNe (2019)} \\ \hline
\textbf{accuracy}                              & 0.49536     & 0.58369      & 0.52423     & 0.54012     & 0.56541      & 0.55255       & 0.50142            & 0.56513             & 0.52652            & 0.53735            & 0.54633             & 0.56500              \\
\textbf{balanced\_accuracy}                    & 0.49260     & 0.50000      & 0.52912     & 0.51277     & 0.49154      & 0.50998       & 0.49117            & 0.50000             & 0.53969            & 0.52019            & 0.50526             & 0.55142              \\
\textbf{average\_precision}                    & 0.63143     & 0.58369      & 0.66616     & 0.65376     & 0.64830      & 0.64028       & 0.62777            & 0.56513             & 0.66468            & 0.64958            & 0.64701             & 0.67791              \\
\textbf{brier\_score}                     & 0.25038     & 0.41631      & 0.44585     & 0.26040     & 0.24539      & 0.26211       & 0.24989            & 0.43487             & 0.42387            & 0.25578            & 0.24803             & 0.24681              \\
\textbf{f1}                                    & 0.45335     & 0.72279      & 0.48915     & 0.57451     & 0.68523      & 0.63161       & 0.40744            & 0.71096             & 0.36850            & 0.51776            & 0.68411             & 0.62999              \\
\textbf{log\_loss}                        & 0.69391     & 14.37905     & 4.96686     & 0.72145     & 0.68406      & 0.72951       & 0.69292            & 15.02025            & 3.80885            & 0.70676            & 0.68920             & 0.69064              \\
\textbf{precision}                             & 0.48834     & 0.58369      & 0.61343     & 0.58415     & 0.57564      & 0.58449       & 0.48552            & 0.56513             & 0.64769            & 0.54002            & 0.57234             & 0.59516              \\
\textbf{recall}                                & 0.47464     & 1.00000      & 0.48868     & 0.65708     & 0.90818      & 0.74119       & 0.38639            & 1.00000             & 0.30030            & 0.56724            & 0.92475             & 0.72064              \\
\textbf{roc\_auc}                              & 0.48174     & 0.50000      & 0.53271     & 0.50914     & 0.50410      & 0.51576       & 0.50932            & 0.50000             & 0.55709            & 0.52027            & 0.53747             & 0.59038              \\
\textbf{accuracy\_weighted}                    & 0.48987     & 0.55932      & 0.53146     & 0.54580     & 0.55537      & 0.54087       & 0.48467            & 0.55554             & 0.55555            & 0.54265            & 0.55042             & 0.55781              \\
\textbf{balanced\_accuracy\_weighted}          & 0.49918     & 0.50000      & 0.54488     & 0.52744     & 0.49988      & 0.50218       & 0.46705            & 0.50000             & 0.54822            & 0.53294            & 0.52086             & 0.54170              \\
\textbf{average\_precision\_weighted}          & 0.63575     & 0.55932      & 0.68716     & 0.67377     & 0.66033      & 0.64071       & 0.64644            & 0.55554             & 0.69751            & 0.70454            & 0.69596             & 0.68794              \\
\textbf{brier\_score\_weighted}           & 0.25055     & 0.44068      & 0.44199     & 0.25940     & 0.24767      & 0.26597       & 0.25018            & 0.44446             & 0.39657            & 0.24981            & 0.24761             & 0.25543              \\
\textbf{f1\_weighted}                          & 0.43849     & 0.69064      & 0.48659     & 0.56997     & 0.66257      & 0.60400       & 0.38017            & 0.69667             & 0.38291            & 0.51829            & 0.67985             & 0.61122              \\
\textbf{log\_loss\_weighted}              & 0.69425     & 15.22079     & 4.88732     & 0.72007     & 0.68868      & 0.73764       & 0.69352            & 15.35149            & 3.53302            & 0.69418            & 0.68828             & 0.71678              \\
\textbf{precision\_weighted}                   & 0.47010     & 0.55932      & 0.61993     & 0.58330     & 0.56094      & 0.56041       & 0.45313            & 0.55554             & 0.68509            & 0.54863            & 0.57665             & 0.58761              \\
\textbf{recall\_weighted}                      & 0.48582     & 1.00000      & 0.51544     & 0.68242     & 0.91340      & 0.73906       & 0.35946            & 1.00000             & 0.30880            & 0.57750            & 0.92763             & 0.70145              \\
\textbf{roc\_auc\_weighted}                    & 0.49189     & 0.50000      & 0.55564     & 0.53794     & 0.51912      & 0.51494       & 0.50356            & 0.50000             & 0.56504            & 0.56866            & 0.56714             & 0.57509              \\
\textbf{avg\_profit}                 & -0.93097    & -0.02457     & 0.88124     & 0.69087     & -0.09201     & 0.41763       & -2.77925           & 0.13442             & 2.53718            & 2.42230            & 0.01231             & 2.97753              \\
\textbf{tot\_profit}                 & -18.15649   & 0.67272      & 16.94377    & 12.37207    & -1.56365     & 9.26234       & -54.55260          & 6.87324             & 48.74426           & 46.00116           & 2.48023             & 61.88742             \\
\textbf{avg\_trading\_profit}        & -0.84473    & -0.02134     & 3.37902     & 2.29790     & -0.04317     & 0.44433       & -2.95184           & 0.13442             & 7.05829            & 5.12391            & 1.33662             & 5.03380              \\
\textbf{std\_trading\_profit}        & 26.59663    & 27.10337     & 20.91999    & 24.13176    & 26.68021     & 25.18487      & 45.71519           & 43.01486            & 25.69383           & 36.64222           & 42.02958            & 37.64435             \\
\textbf{downw\_std\_trading\_profit} & 14.35936    & 16.97141     & 10.19751    & 13.70709    & 16.33821     & 15.07755      & 29.52038           & 28.98308            & 10.34339           & 22.77312           & 27.86772            & 23.78301             \\
\textbf{avg\_trades}                 & 0.47470     & 1.00000      & 0.45360     & 0.63924     & 0.91416      & 0.73737       & 0.39335            & 1.00000             & 0.25023            & 0.54021            & 0.91661             & 0.68025              \\ \hline
\end{tabular}
\caption{mean of the metrics on the test sets for various models in experiment 1.1}
\end{table}

\newpage

\section{Metrics of experiment 1.2}

\begin{table}[htp]
\tiny
\begin{tabular}{lllllllllllll}
\hline
\textbf{}                                           & \textbf{AB} & \textbf{All} & \textbf{GB} & \textbf{RF} & \textbf{SVC} & \textbf{kNNe} & \textbf{AB (2019)} & \textbf{All (2019)} & \textbf{GB (2019)} & \textbf{RF (2019)} & \textbf{SVC (2019)} & \textbf{kNNe (2019)} \\ \hline
\textbf{accuracy}                              & 0.51144     & 0.58283      & 0.52775     & 0.53713     & 0.57395      & 0.54679       & 0.52040            & 0.56126             & 0.49779            & 0.52129            & 0.55662             & 0.57615              \\
\textbf{balanced\_accuracy}                    & 0.49304     & 0.50000      & 0.51308     & 0.51352     & 0.49943      & 0.50895       & 0.51225            & 0.50000             & 0.52109            & 0.50356            & 0.49514             & 0.54525              \\
\textbf{average\_precision}                    & 0.61690     & 0.58283      & 0.62035     & 0.62745     & 0.62443      & 0.61231       & 0.60652            & 0.56126             & 0.59843            & 0.59521            & 0.63409             & 0.62812              \\
\textbf{brier\_score}                     & 0.24990     & 0.41717      & 0.44507     & 0.25811     & 0.24306      & 0.26742       & 0.24986            & 0.43874             & 0.47023            & 0.25640            & 0.24547             & 0.25010              \\
\textbf{f1}                                    & 0.47971     & 0.73300      & 0.54222     & 0.61640     & 0.70978      & 0.64769       & 0.43184            & 0.71654             & 0.36346            & 0.61617            & 0.70952             & 0.66768              \\
\textbf{log\_loss}                        & 0.69294     & 14.40879     & 4.92848     & 0.71642     & 0.67920      & 0.90804       & 0.69287            & 15.15375            & 4.51528            & 0.70675            & 0.68405             & 0.69637              \\
\textbf{precision}                             & 0.54744     & 0.58283      & 0.59261     & 0.60153     & 0.58421      & 0.58924       & 0.65610            & 0.56126             & 0.59037            & 0.57036            & 0.55886             & 0.59574              \\
\textbf{recall}                                & 0.51699     & 1.00000      & 0.54777     & 0.67968     & 0.92783      & 0.74075       & 0.40141            & 1.00000             & 0.26837            & 0.69840            & 0.97847             & 0.76807              \\
\textbf{roc\_auc}                              & 0.50405     & 0.50000      & 0.51659     & 0.52334     & 0.52113      & 0.50796       & 0.50408            & 0.50000             & 0.51885            & 0.50895            & 0.55882             & 0.56174              \\
\textbf{accuracy\_weighted}                    & -37.45949   & 0.53263      & 0.52074     & -29.57363   & -20.63826    & -25.42804     & 0.94798            & 0.53142             & 0.51110            & 0.07104            & 0.79786             & 0.29726              \\
\textbf{balanced\_accuracy\_weighted}          & 0.48143     & 0.50000      & 0.50886     & 0.51959     & 0.49842      & 0.49761       & 0.48800            & 0.50000             & 0.51951            & 0.54637            & 0.51483             & 0.56022              \\
\textbf{average\_precision\_weighted}          & 0.59357     & 0.53263      & 0.59275     & 0.60839     & 0.59878      & 0.56608       & 0.57246            & 0.53142             & 0.60526            & 0.63117            & 0.67746             & 0.61027              \\
\textbf{brier\_score\_weighted}           & 0.24987     & 0.46737      & 0.45270     & 0.26370     & 0.25101      & 0.27782       & 0.25022            & 0.46858             & 0.45811            & 0.24973            & 0.24402             & 0.25190              \\
\textbf{f1\_weighted}                          & 2.04778     & 0.68546      & 0.52782     & 8.28274     & 9.55666      & 2.34614       & 0.75624            & 0.68808             & 0.37310            & 1.29745            & 2.60897             & 1.32178              \\
\textbf{log\_loss\_weighted}              & 0.69289     & 16.14269     & 5.09233     & 0.72852     & 0.69538      & 0.92489       & 0.69358            & 16.18466            & 4.42845            & 0.69237            & 0.68104             & 0.69869              \\
\textbf{precision\_weighted}                   & -36.04751   & 0.53263      & 0.55495     & -31.26805   & -14.27970    & -31.13146     & -12.12129          & 0.53142             & 0.60243            & 5.47984            & 0.50370             & 1.57704              \\
\textbf{recall\_weighted}                      & 0.52051     & 1.00000      & 0.56625     & 0.68586     & 0.92569      & 0.72352       & 0.39704            & 1.00000             & 0.27313            & 0.68083            & 0.97466             & 0.76129              \\
\textbf{roc\_auc\_weighted}                    & 0.49558     & 0.50000      & 0.51568     & 0.52667     & 0.51619      & 0.48540       & 0.45473            & 0.50000             & 0.53632            & 0.56754            & 0.61121             & 0.55606              \\
\textbf{avg\_profit}                 & -0.19468    & -0.07855     & 0.32966     & 0.35145     & 0.15763      & 0.00363       & 0.42947            & 0.01312             & 1.90077            & 1.92852            & 1.12093             & 3.23932              \\
\textbf{tot\_profit}                 & -8.01564    & 1.83717      & 25.90335    & 23.72656    & 14.21126     & 6.44040       & 20.92325           & 1.12517             & 117.53313          & 118.61530          & 68.29716            & 202.79838            \\
\textbf{avg\_trading\_profit}        & 1.52385     & -0.07855     & 1.54310     & 1.17640     & 0.22195      & -0.11392      & 15.58090           & 0.01312             & 9.81098            & 2.95481            & 1.11367             & 4.39896              \\
\textbf{std\_trading\_profit}        & 28.56531    & 31.51894     & 31.26535    & 29.99389    & 30.85032     & 30.79933      & 37.85243           & 52.45532            & 48.79117           & 46.96594           & 49.52913            & 49.12192             \\
\textbf{downw\_std\_trading\_profit} & 22.01835    & 25.01891     & 20.74342    & 23.21967    & 24.07020     & 24.35686      & 36.41063           & 44.69394            & 29.28331           & 40.68800           & 40.46170            & 41.08577             \\
\textbf{avg\_trades}                 & 0.52252     & 1.00000      & 0.54166     & 0.66846     & 0.92678      & 0.73543       & 0.39394            & 1.00000             & 0.25212            & 0.69706            & 0.98229             & 0.72845              \\ \hline
\end{tabular}
\caption{mean of the metrics on the test sets for various models in experiment 1.2}
\end{table}

\newpage

\section{Metrics of experiment 2.1}

\begin{table}[htp]
\tiny
\begin{tabular}{@{}lllllllllllll@{}}
\toprule
\textbf{}                                           & \textbf{AB} & \textbf{All} & \textbf{GB} & \textbf{RF} & \textbf{SVC} & \textbf{kNNe} & \textbf{AB (2019)} & \textbf{All (2019)} & \textbf{GB (2019)} & \textbf{RF (2019)} & \textbf{SVC (2019)} & \textbf{kNNe (2019)} \\ \midrule
\textbf{accuracy}                              & 0.52925     & 0.58466      & 0.52445     & 0.51458     & 0.57344      & 0.53932       & 0.50871             & 0.56126              & 0.48306             & 0.51743             & 0.55989              & 0.55878               \\
\textbf{balanced\_accuracy}                    & 0.50091     & 0.50000      & 0.51466     & 0.49156     & 0.49786      & 0.50239       & 0.50083             & 0.50000              & 0.50313             & 0.51652             & 0.50418              & 0.53545               \\
\textbf{average\_precision}                    & 0.62116     & 0.58466      & 0.61781     & 0.60914     & 0.61308      & 0.59426       & 0.61352             & 0.56126              & 0.59249             & 0.60040             & 0.64036              & 0.61663               \\
\textbf{brier\_score}                     & 0.24931     & 0.41534      & 0.44503     & 0.26617     & 0.24294      & 0.27245       & 0.24969             & 0.43874              & 0.47772             & 0.25794             & 0.24551              & 0.25247               \\
\textbf{f1}                                    & 0.50389     & 0.73466      & 0.51528     & 0.53765     & 0.70652      & 0.62870       & 0.40030             & 0.71654              & 0.28324             & 0.55599             & 0.69678              & 0.65042               \\
\textbf{log\_loss}                        & 0.69176     & 14.34577     & 4.92201     & 0.74011     & 0.67898      & 1.07025       & 0.69252             & 15.15375             & 4.93607             & 0.70996             & 0.68413              & 0.70189               \\
\textbf{precision}                             & 0.56135     & 0.58466      & 0.57748     & 0.58157     & 0.58309      & 0.58067       & 0.49498             & 0.56126              & 0.48288             & 0.58280             & 0.56463              & 0.58667               \\
\textbf{recall}                                & 0.55719     & 1.00000      & 0.51353     & 0.57221     & 0.91533      & 0.71300       & 0.37700             & 1.00000              & 0.21792             & 0.57643             & 0.91951              & 0.74409               \\
\textbf{roc\_auc}                              & 0.51380     & 0.50000      & 0.52080     & 0.50214     & 0.51350      & 0.49047       & 0.51307             & 0.50000              & 0.51160             & 0.51757             & 0.56156              & 0.56313               \\
\textbf{accuracy\_weighted}                    & 0.51874     & 0.53667      & 0.53426     & 0.51193     & 0.54171      & 0.51682       & 0.48558             & 0.53142              & 0.50659             & 0.53198             & 0.57844              & 0.55376               \\
\textbf{balanced\_accuracy\_weighted}          & 0.49578     & 0.50000      & 0.52707     & 0.49566     & 0.50282      & 0.49414       & 0.47182             & 0.50000              & 0.50408             & 0.54016             & 0.54708              & 0.54205               \\
\textbf{average\_precision\_weighted}          & 0.61358     & 0.53667      & 0.60605     & 0.58548     & 0.58502      & 0.55098       & 0.62079             & 0.53142              & 0.62385             & 0.62137             & 0.67948              & 0.59784               \\
\textbf{brier\_score\_weighted}           & 0.24947     & 0.46333      & 0.44013     & 0.26964     & 0.24956      & 0.28390       & 0.24997             & 0.46858              & 0.46246             & 0.25286             & 0.24278              & 0.25280               \\
\textbf{f1\_weighted}                          & 0.47755     & 0.68864      & 0.49824     & 0.52146     & 0.66780      & 0.58292       & 0.36235             & 0.68808              & 0.24723             & 0.53726             & 0.69462              & 0.62813               \\
\textbf{log\_loss\_weighted}              & 0.69208     & 16.00322     & 4.84008     & 0.75049     & 0.69244      & 1.03708       & 0.69309             & 16.18466             & 4.97101             & 0.69890             & 0.67856              & 0.70148               \\
\textbf{precision\_weighted}                   & 0.51939     & 0.53667      & 0.54618     & 0.54724     & 0.53919      & 0.52594       & 0.43629             & 0.53142              & 0.47880             & 0.56405             & 0.56920              & 0.56179               \\
\textbf{recall\_weighted}                      & 0.55577     & 1.00000      & 0.52743     & 0.58653     & 0.91588      & 0.69027       & 0.34499             & 1.00000              & 0.18782             & 0.57296             & 0.91671              & 0.74079               \\
\textbf{roc\_auc\_weighted}                    & 0.52106     & 0.50000      & 0.54433     & 0.50006     & 0.51067      & 0.47461       & 0.50032             & 0.50000              & 0.56517             & 0.54939             & 0.61799              & 0.55441               \\
\textbf{avg\_profit}                 & 0.17085     & 0.00662      & 0.48056     & 0.12148     & 0.42232      & -0.00923      & 0.21254             & 0.01312              & 1.06800             & 1.80883             & 2.08823              & 1.97236               \\
\textbf{tot\_profit}                 & 16.70874    & 6.69076      & 34.47599    & 12.92784    & 32.16957     & 3.71116       & 7.29327             & 1.12517              & 65.91010            & 110.04686           & 127.80742            & 120.77171             \\
\textbf{avg\_trading\_profit}        & -1.30998    & 0.00662      & 0.64614     & 0.10826     & 0.29036      & -0.37553      & 1.26337             & 0.01312              & 0.31608             & 2.95859             & 2.44030              & 2.74972               \\
\textbf{std\_trading\_profit}        & 27.34241    & 31.22115     & 23.85985    & 30.78238    & 30.24300     & 30.14789      & 43.95953            & 52.45532             & 20.56710            & 48.13272            & 47.45548             & 49.83072              \\
\textbf{downw\_std\_trading\_profit} & 20.79549    & 24.60179     & 16.68981    & 22.98998    & 22.85209     & 23.83082      & 35.61971            & 44.69394             & 13.81221            & 41.97128            & 36.92310             & 41.10563              \\
\textbf{avg\_trades}                 & 0.56112     & 1.00000      & 0.50679     & 0.58061     & 0.91657      & 0.71274       & 0.37880             & 1.00000              & 0.21801             & 0.56112             & 0.91546              & 0.71475               \\ \bottomrule
\end{tabular}
\caption{mean of the metrics on the test sets for various models in experiment 2.1}
\end{table}

\newpage

\section{Metrics of experiment 2.2}
\begin{table}[htp]
\tiny
\begin{tabular}{lllllllllllll}
\hline
\textbf{}                                           & \textbf{AB} & \textbf{All} & \textbf{GB} & \textbf{RF} & \textbf{SVC} & \textbf{kNNe} & \textbf{AB (2019)} & \textbf{All (2019)} & \textbf{GB (2019)} & \textbf{RF (2019)} & \textbf{SVC (2019)} & \textbf{kNNe (2019)} \\ \hline
\textbf{accuracy}                              & 0.50292     & 0.58457      & 0.51827     & 0.52382     & 0.57229      & 0.54210       & 0.50729             & 0.56513              & 0.55042             & 0.54769             & 0.53917              & 0.57669               \\
\textbf{balanced\_accuracy}                    & 0.50728     & 0.50000      & 0.51691     & 0.50442     & 0.50125      & 0.50540       & 0.50802             & 0.50000              & 0.55346             & 0.53258             & 0.50353              & 0.57753               \\
\textbf{average\_precision}                    & 0.64956     & 0.58457      & 0.65849     & 0.64209     & 0.63654      & 0.63733       & 0.65376             & 0.56513              & 0.67506             & 0.65610             & 0.64847              & 0.66300               \\
\textbf{brier\_score}                     & 0.25010     & 0.41543      & 0.45188     & 0.27017     & 0.24562      & 0.26600       & 0.24985             & 0.43487              & 0.41311             & 0.26164             & 0.24808              & 0.25112               \\
\textbf{f1}                                    & 0.47889     & 0.72440      & 0.48158     & 0.49376     & 0.68382      & 0.62340       & 0.42947             & 0.71096              & 0.41219             & 0.41876             & 0.64611              & 0.63585               \\
\textbf{log\_loss}                        & 0.69334     & 14.34867     & 4.96021     & 0.74500     & 0.68460      & 0.75480       & 0.69285             & 15.02025             & 3.57815             & 0.72242             & 0.68929              & 0.69974               \\
\textbf{precision}                             & 0.54180     & 0.58457      & 0.56863     & 0.53183     & 0.58238      & 0.59007       & 0.51290             & 0.56513              & 0.59481             & 0.51942             & 0.56328              & 0.62260               \\
\textbf{recall}                                & 0.51003     & 1.00000      & 0.48343     & 0.55919     & 0.90194      & 0.72806       & 0.41863             & 1.00000              & 0.35112             & 0.43548             & 0.85673              & 0.72560               \\
\textbf{roc\_auc}                              & 0.49442     & 0.50000      & 0.52978     & 0.48755     & 0.48213      & 0.51089       & 0.53521             & 0.50000              & 0.58122             & 0.54486             & 0.53075              & 0.58991               \\
\textbf{accuracy\_weighted}                    & 0.49582     & 0.56166      & 0.52116     & 0.52796     & 0.56705      & 0.52700       & 0.49856             & 0.55554              & 0.54955             & 0.57856             & 0.53964              & 0.55766               \\
\textbf{balanced\_accuracy\_weighted}          & 0.50756     & 0.50000      & 0.52818     & 0.52144     & 0.50933      & 0.50102       & 0.49680             & 0.50000              & 0.54354             & 0.57063             & 0.51258              & 0.56739               \\
\textbf{average\_precision\_weighted}          & 0.65374     & 0.56166      & 0.66977     & 0.65533     & 0.64807      & 0.64132       & 0.67506             & 0.55554              & 0.68962             & 0.68196             & 0.68910              & 0.67374               \\
\textbf{brier\_score\_weighted}           & 0.25017     & 0.43834      & 0.44885     & 0.26870     & 0.24671      & 0.26754       & 0.25003             & 0.44446              & 0.41078             & 0.25832             & 0.24741              & 0.25474               \\
\textbf{f1\_weighted}                          & 0.46372     & 0.69289      & 0.47783     & 0.48967     & 0.66249      & 0.59203       & 0.41756             & 0.69667              & 0.40196             & 0.44394             & 0.63595              & 0.62328               \\
\textbf{log\_loss\_weighted}              & 0.69348     & 15.14007     & 4.81033     & 0.74257     & 0.68673      & 0.74123       & 0.69321             & 15.35149             & 3.44844             & 0.71628             & 0.68787              & 0.70613               \\
\textbf{precision\_weighted}                   & 0.53089     & 0.56166      & 0.56228     & 0.52703     & 0.57056      & 0.56420       & 0.50819             & 0.55554              & 0.59172             & 0.54526             & 0.57111              & 0.60222               \\
\textbf{recall\_weighted}                      & 0.51326     & 1.00000      & 0.51461     & 0.58889     & 0.90566      & 0.72905       & 0.40401             & 1.00000              & 0.35211             & 0.47559             & 0.85761              & 0.74317               \\
\textbf{roc\_auc\_weighted}                    & 0.50039     & 0.50000      & 0.55943     & 0.50949     & 0.50115      & 0.52263       & 0.52923             & 0.50000              & 0.58954             & 0.56302             & 0.55201              & 0.58386               \\
\textbf{avg\_profit}                 & -0.62820    & 0.05483      & 0.97600     & 0.87244     & 0.25511      & 0.09110       & -2.02822            & 0.13442              & 2.82515             & 3.85740             & -0.17795             & 1.74368               \\
\textbf{tot\_profit}                 & -11.26972   & 2.24856      & 20.17385    & 16.01987    & 4.92787      & 2.85817       & -39.36596           & 6.87324              & 56.36627            & 77.80045            & -1.70800             & 37.82217              \\
\textbf{avg\_trading\_profit}        & 0.95750     & 0.05483      & 2.53667     & 2.52749     & 0.55207      & 0.43967       & -1.11188            & 0.13442              & 7.69372             & 11.37158            & 1.37099              & 4.09908               \\
\textbf{std\_trading\_profit}        & 25.18052    & 26.93153     & 23.22397    & 22.18854    & 26.09097     & 25.45523      & 43.34379            & 43.01486             & 34.39908            & 32.30414            & 41.46429             & 40.41335              \\
\textbf{downw\_std\_trading\_profit} & 14.72153    & 16.99380     & 11.92305    & 13.44537    & 15.75349     & 14.23827      & 29.43013            & 28.98308             & 22.19952            & 21.44452            & 27.34706             & 25.71597              \\
\textbf{avg\_trades}                 & 0.49407     & 1.00000      & 0.45779     & 0.54556     & 0.89912      & 0.71682       & 0.40679             & 1.00000              & 0.29782             & 0.39625             & 0.84891              & 0.66109               \\ \hline
\end{tabular}
\caption{mean of the metrics on the test sets for various models in experiment 2.2}
\end{table}

\end{landscape}
\end{document}